\documentclass[prl,twocolumn, showpacs]{revtex4}
\usepackage{graphicx}
\usepackage{graphics}
\usepackage{amssymb}
\usepackage{epstopdf}
\usepackage{color}
\usepackage{subfigure}
\usepackage{epsfig}

\begin{document}
\title{Detecting Entanglement with Jarzynski's Equality}
\author{Jenny Hide$^1$ and Vlatko Vedral$^{2,3,4}$}
\affiliation{Department of Physics and Astronomy, EC Stoner Building, University of Leeds, Leeds, LS2 9JT, UK$^1$\\
Clarendon Laboratory, University of Oxford, Parks Road, Oxford OX1 3PU, United Kingdom$^2$\\
Centre for Quantum Technologies, National University of Singapore, 3 Science Drive 2, Singapore 117543$^3$ \\
Department of Physics, National University of Singapore, 2 Science Drive 3, Singapore 117542$^4$.}

\date{\today}

\begin{abstract}

We present a method for detecting the entanglement of a state
using non-equilibrium processes. A comparison of relative entropies
allows us to construct an entanglement witness. The relative entropy
can further be related to the quantum
Jarzynski equality, allowing non-equilibrium work to be used in entanglement
detection. To exemplify our results, we consider two different spin chains.

\end{abstract}

\maketitle


In quantum information theory, entanglement is considered
not only an interesting phenomenon, but also a resource which can be
used in quantum computation. Entanglement has therefore been the
topic of much research. A separable state can be written as a
convex sum of pure product states, $\sigma=\sum_i p_i
\sigma_i^1 \otimes \sigma_i^2 \otimes \cdots \otimes \sigma_i^n$
where the $p_i$s are the weights of the product states, $\sigma_i^j$,
with $\sum_ip_i =1$, while an entangled state cannot. Many methods have
been devised to measure
and detect entanglement, even for thermal and for many-body systems
\cite{manybody}. The entanglement witness \cite{ent_wit} is 
an expectation value of an operator which is bounded for any separable
state, whereas entangled states can exceed this bound.
A thermodynamic witness allows us to use thermodynamic quantities
such as the magnetic susceptibility \cite{mag_susc} to detect entanglement.
The major advantage of using a such a witness is that we can detect thermal
many-body entanglement using experimentally measurable quantities.

Thus far, these thermodynamic witnesses have only been used for
detecting entanglement in equilibrium systems. However, a result from
condensed matter theory, Jarzynski's equality \cite{Jarzynski}, allows
the change in free energy between two equilibrium states to be related to
the non-equilibrium work done needed to drive the system from one state to the
other. While the work done can be measured or calculated in an experiment,
the change in free energy cannot. Thus Jarzynski's equality can be used
to experimentally estimate the change in free energy during a non-equilibrium
process \cite{ExpJar}. It is the aim of this letter to use Jarzynski's equality
to witness equilibrium entanglement using non-equilibrium processes.
Our work also raises the exciting possibility of using this witness to detect
entanglement in biological systems.


In our construction of an entanglement witness, we use the relative
entropy, a directed distance from an initial state $\rho_i$ to
a final state $\rho_f$, given by

\begin{equation}
S(\rho_f ||\rho_i ) = tr(\rho_f \log \rho_f) - tr(\rho_f \log \rho_i).
\label{eq:RelEntEq}
\end{equation}

\noindent This is a \emph{measure} of entanglement \cite{VrelEnt} when
$\rho_i=\sigma_{css}$ is the closest separable state to $\rho_f$. Formally,
we have for the relative entropy of entanglement,
$E_{RE}(\rho_f) = \min_{\rho_i \in {\mathcal S}} S(\rho_f || \rho_i )$,
where we take the minimum over the set of separable states
${\mathcal S}$, to find $\sigma_{css}$. The relative entropy can measure
entanglement for both equilibrium and non-equilibrium, pure and mixed states,
and therefore for thermal, open and closed systems.

\begin{figure}[t]
\begin{center}
\centerline{
\includegraphics[width=3.0in]{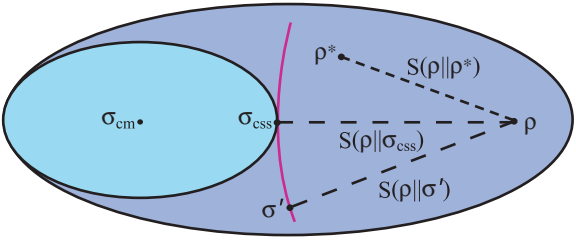} }
\end{center}
\caption{This is a $2$d representation of the multidimensional
Hilbert space. The small oval represents the set of separable
states, and the large oval the set of all states.
$\sigma_{cm}$ is a completely mixed state, and $\sigma_{css}$ is the closest
separable state to $\rho$. Any state along the pink curve, such as $\sigma'$,
has the same ``distance" in terms of the relative entropy as $\sigma_{css}$
to $\rho$.}
\label{fig1}
\end{figure}

We can now construct an entanglement witness using the relative entropy by
introducing an arbitrary state $\rho^*$. Since the set of separable states
is convex, and the relative entropy
is a directed distance, if the distance from $\sigma_{css}$ to $\rho$
is larger than the distance from $\rho^*$ to $\rho$,
then $\rho^*$ is entangled. Fig. (\ref{fig1}) gives a
two dimensional representation of this idea. Hence our witness is

\begin{equation}
S(\rho || \sigma_{css}) \geq S(\rho || \rho^*).
\label{eq:RelE_ineq}
\end{equation}

\noindent If $\rho^*$ satisfies this inequality, we know it must be
entangled. The witness is best when $\rho$ is a pure state
and hence is located at the edge of the outer ellipse in Fig. (\ref{fig1}).
We will refer to this inequality as the relative entropy witness.
We note that this witness can detect entanglement in both equilibrium
and non-equilibrium states.


Although originally a
classical result, it has been shown that Jarzynski's equality,

\begin{equation}
\langle e^{-\beta {\mathcal W}} \rangle = e^{-\beta \Delta F},
\label{jarzy}
\end{equation}

\noindent where $\beta^{-1}$ is the temperature, ${\mathcal W}$ is the
work done on the system and $\Delta F$ is the change in free energy between
the initial and final equilibrium states, is valid for both open \cite{openJnew}
and closed quantum systems \cite{Kurchan2,Tasaki,Mukamel}. The brackets
$\langle \cdots \rangle$ denote an average
over all possible realisations of the work, or trajectories in phase space.
Both the path and the rate at which the system is driven are fixed for the
equality, though each are arbitrary.

There are several different methods (for a review, see reference
\cite{Kurchan_rev}), in the literature for deriving a quantum
version of Jarzynski's equality, however we discuss the one which has been
successfully theoretically verified \cite{Engel}.
In a closed quantum system, instead of classical trajectories in phase
space, we define the quantum equivalent of quantum transition probabilities
\cite{Kurchan2,Tasaki,Mukamel}.
An initial Hamiltonian $H_i$ and a final Hamiltonian $H_f$
have eigenvalues $E_n^i$, $E_m^f$ and eigenvectors $| \phi_n^i \rangle$,
$| \phi_m^f \rangle$ respectively.
We perform a measurement of the energy at time $t_i$ and then again at $t_f$ so
that the system
is in a specific energy eigenstate. The quantum transition probabilities are
then defined as $q_{m,n}=|\langle \phi_m^f  | U(t_f)| \phi_n^i \rangle |^2$
where $U(t_f)=\hat{T}_<e^{-i\int_0^{t_f} H(s) ds}$ is the time
evolution operator, and $\hat{T}_<$ is the time ordering operator. $q_{m,n}$
can be interpreted as the probability
that the final state of the system is $| \phi_m^f \rangle$ given that it was
initially in the state $| \phi_n^i \rangle$.
The average is then given as $\langle e^{-\beta {\mathcal W}} \rangle = \sum_{n}
(e^{-\beta E_n^i}/Z_i) \sum_{m} q_{m,n} e^{-\beta(E_m^f-E_n^i)} $ where the work is
defined as ${\mathcal W}=E_m^f-E_n^i$ and $Z_i=tr(e^{-\beta H_i})$ is the initial
partition function.

Consider now an open quantum system (subsystem, $S$) interacting
with a bath, $B$, with total Hamiltonian $H(t)=H_S(t)+H_{SB}+H_B$ and
arbitrary coupling, $H_{SB}$ \cite{openJnew}. As
only the subsystem is time dependent, the change in energy of the
total system equals the work done on the subsystem. Hence the average
in equation (\ref{jarzy}) is identical to the closed system case.
Further, the free energy of the total system
is given by $F(t) = F_S(t) + F_B$. This allows Jarzynski's equality
to be written $\langle e^{-\beta {\mathcal W}} \rangle = e^{-\beta
\Delta F_S}$ \cite{openJnew}.

Other fluctuation theorems have also been derived. One equality
which will be useful \cite{Tasaki} is $\langle e^{-(\beta_f E_f
- \beta_i E_i) } \rangle = e^{-(\beta_f F_f- \beta_i F_i )}$. This demonstrates
that a change in temperature between the initial and final state can also be
taken into account. However, unless $\beta_i=\beta_f$, the quantity $(\beta_f E_f
- \beta_i E_i)$ no longer relates to work. We refer to this equation as the
Jarzynski-Tasaki equality.


Consider again the relative entropy, equation (\ref{eq:RelEntEq}).
We now restrict each of the states to be in thermal equilibrium. Thus
we have initial and final states,
$\sigma_{css}=e^{-\beta_i H_i}/Z_i$ and $\rho=e^{-\beta_f H_f}/Z_f$
respectively. Similarly,
$\rho^*=e^{-\beta^* H^*}/Z^*$. Expanding equation (\ref{eq:RelEntEq}), we
can write the relative entropy in terms of a change in free energy
\cite{donald,vlatko_land}, $S(\rho || \sigma_{css} ) =
\Delta (\beta F) - tr( \rho_f \Delta (\beta H) )$
where $\Delta (\beta F) = \beta_f F_f-\beta_i F_i$
and $\Delta (\beta H) =  \beta_f H_f-\beta_i H_i$.
Combining this identity with the Jarzynski-Tasaki equality, we find that

\begin{equation}
S(\rho || \sigma_{css} ) = - tr( \rho \Delta (\beta H) ) -
\ln \langle e^{-(\beta_f E_f - \beta_i E_i)} \rangle.
\label{eq:relentJ}
\end{equation}

\noindent Equation (\ref{eq:relentJ}) relates the
entanglement to the average change in energy at different temperatures
(in a possibly driven system). When $ \beta_f= \beta_i$, we can
instead relate the entanglement to the average work done
in creating the quantum correlations of $\rho$ from
the purely classical correlations of $\sigma_{css}$.
In addition to this being an interesting result in itself, we
can also use this definition of the relative entropy in the
entanglement witness, equation (\ref{eq:RelE_ineq}). We call this the
relative Jarzynski witness and discuss this in more detail below.


An open quantum system (or subsystem) can also be considered.
As discussed previously, such systems
also obey Jarzynski's equality, $\langle e^{-\beta {\mathcal W}} \rangle =
e^{-\beta \Delta F_S}$ where $F_S(t)$ is the free energy of the subsystem. We define
$Y(t)=tr(e^{-\beta (H_S(t)+H_{SB}+H_B)})$ as the partition function
of the total system and $Z_B=tr(e^{-\beta H_B})$ as the
partition function of the bath. The partition function of the
subsystem, $Z_S(t)=tr(e^{-\beta F_S})=Y(t)/Z_B$ can be associated with an
effective Hamiltonian \cite{openJnew},

\begin{equation}
H^{eff}(t) = -\frac{1}{\beta} \ln \left[ \frac{tr_B(e^{-\beta (H_S(t)+
H_{SB}+H_B)})}{tr_B(e^{-\beta H_B})} \right],
\end{equation}

\noindent so that $Z_S(t)=tr_S(e^{-\beta H^{eff}(t)})$. Using these equations,
and since the initial and final states must be in equilibrium, we have
$\rho_S= e^{-\beta H^{eff}(t)}/Z_S(t)$ for each state. This only
represents the state of the system when it is in equilibrium.

The relative entropy can now be defined in terms of the
effective Hamiltonian and the work done on the subsystem,
$S(\rho_{S} || \sigma_{S,css}) = -\beta tr[\rho_{S} (H_f^{eff}-
H_i^{eff})] - \ln \langle e^{-\beta {\mathcal W}} \rangle$.
Hence we can write the entanglement witness for
an open quantum system, $S(\rho_S || \sigma_{S,css}) \geq S(\rho_S ||\rho_S^*)$,
where $\rho_S^*=e^{-\beta H^{eff,*}}/Z_S^*$ as the state is in equilibrium.

We have shown that it is possible to detect entanglement in a state
$\rho^*$ or $\rho_S^*$ using a non-equilibrium process.
There are three ways we can use the entanglement witness. First, if
we have a specific state $\rho^*$ in mind but don't know whether
it is entangled, the witness allows us to detect
entanglement in this state. Second, if we have the Hamiltonian of
a system, we can detect entanglement in that system.  
We ask for which values of the parameters of the system, such as a magnetic 
field, is the system entangled? In this case we find out in which $\rho^*$s 
of the system entanglement can be detected. 
Computationally, linking equation (\ref{eq:RelE_ineq}) to Jarzynski's equality
simply gives a different way to calculate
the relative entropy. It is in the third method, the experimental applications, that
are exciting in this respect.

Experimentally, we can relate the relative Jarzynski witness to
non-equilibrium processes.
For the closed quantum system when $\beta_f= \beta_i$, and the open quantum
system, this corresponds to a series of measurements. We drop the subscript
$S$ that denotes the subsystem in the open quantum
system here since the work done in the open system is equal to the change
in energy of the total, closed, system. Hence the discussion is valid for
both open and closed systems.

Since we consider a quantum system, measurements of the
energy on many replicas of the same system
will give different values. As the quantum Jarzynski equality
demonstrates, each time we measure an initial and a final energy of a
system to calculate the work, we obtain different results. After many
measurements of the initial state ($\sigma_{css}$) and the final
state ($\rho$), we can calculate the average $\langle e^{-\beta
{\mathcal W}} \rangle$. We then repeat this procedure with
initial state $\rho^*$ and compare the resulting experimental
values of the relative entropy. The value of $tr(\rho (H_f- H_i))$ can
also be experimentally measured. For instance, if a magnetic field
is driving the process, this corresponds to the change in the field
multiplied by the final state magnetisation. We can now detect
entanglement in $\rho^*$.

When $\beta_f \neq \beta_i$ we can use the Jarzynski-Tasaki equality and a similar
argument holds. However, it is no longer the work done that is measured. Instead we
measure the initial and final temperatures of the system in addition to the
energy eigenvalues.

A problem with using our Jarzynski witness is the possibility that
$\sigma_{css}$ is not an equilibrium state of the system, and therefore
we cannot define Jarzynski's equality. However, we find that we do not
require $\sigma_{css}$ to be in equilibrium itself.
Instead, we require only that we have an equilibrium state $\sigma '$
of the system where
$S(\rho || \sigma_{css})=S(\rho || \sigma')$. The states satisfying this
equality are represented by the pink curve in Fig. (\ref{fig1}).


\begin{figure}[t]
\begin{center}
\centerline{
\includegraphics[width=3.0in]{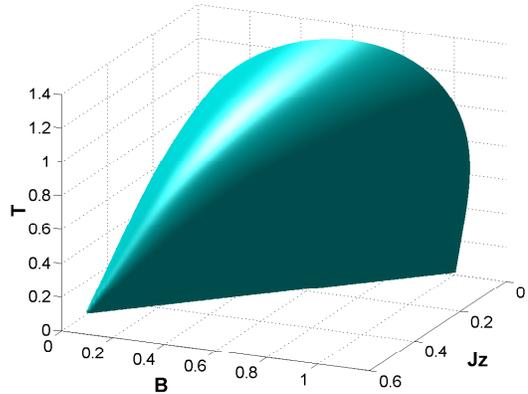} }
\end{center}
\caption{This plots $J_z$ versus $B$ and $T$ when $N=3$, and shows the values
for which we can detect entanglement in $\rho^*$. The state is entangled
between the surface of the plot and the axes.}
\label{fig2}
\end{figure}

We now illustrate the entanglement witness with two examples.
We first consider a three qubit $XXZ$ spin chain as we can define both the
initial and final states to be in equilibrium. In the second example,
we use a seven qubit chain to demonstrate what happens when
the closest separable state is not in equilibrium and we must use a different
equilibrium state $\sigma'$. 

For each example, we have calculated
the witness using the relative entropy witness and using the Jarzynski-Tasaki
witness, and find both give the correct results. This also allows us to successfully
numerically verify the Jarzynski-Taskaki equality. We calculate the time
evolution operator exactly in the three qubit case as $[H(t_1),H(t_2)] = 0$,
and using the method described in \cite{Dorosz}
for seven qubits as $[H(t_1),H(t_2)] \neq 0$. This method allows an approximation of
$U(t_f)$ to be calculated using $U(t_f)=\prod_{n=0}^{M-1} e^{-i H(t_n) \Delta t}$.
We use $\Delta t = 0.001$ to give accurate results. We use these
examples rather than that of an open quantum system since
the closest separable state to $\rho_n$ given below is known. This allows us to
do some of the calculation analytically which allows further insight into the problem.

We take our state to be close to the pure symmetric state,
$\rho_n = (1/n)\hat{S} (|00 \cdots 01 \rangle ) \hat{S} (\langle 00 \cdots 01 | )$ where
$\hat{S}$ is the total symmetrisation operator, whose
closest separable state \cite{vlatko_css,wei_css} is known to be

\begin{equation}
\sigma_{css,n}= \frac{1}{n^n} \sum_{k=0}^n (n-1)^k \hat{S} (|\underbrace{000}_{k}
...\underbrace{111}_{n-k} \rangle ) \hat{S} (\langle \underbrace{000}_{k}...
\underbrace{111}_{n-k} | ).
\label{sigcss}
\end{equation}

\noindent We will identify the states $\rho_n$ and $\sigma_{css,n}$ with
thermal equilibrium states, $e^{-\beta H}/Z$, and hence we will not have exactly the states
above. However, we find that the relative entropy calculated in each case is identical
to many significant figures.

The Hamiltonian of the $XXZ$ spin chain is

\begin{equation}
H=-\sum_{l=1}^n \left[ \frac{J}{2} \left( \sigma_l^x \sigma_{l+1}^x + \sigma_l^y
\sigma_{l+1}^y \right) + J_z\sigma_l^z \sigma_{l+1}^z + B\sigma_l^z \right],
\label{eq:xxz}
\end{equation}

\noindent where $J$ and $J_z$ are coupling strengths, and $B$ is a
magnetic field.

For our first example, the three qubit spin chain, we require
$J_z$ and $B$ to be time dependent. Both $\rho_3$ and $\sigma_{css,3}$
can be written as thermal equilibrium states, $\rho = e^{-\beta H}/Z$, of the
Hamiltonian. For the initial state to be
$\sigma_{css,3}$, we require that $B_i= \beta^{-1} \log(2)/2$
and $J_{z,i}=(2J - \beta^{-1} \log(3))/4$ at a low temperature.
For concreteness, we take $\beta^{-1}=0.01$ and $J=1$.
For the final state to be $\rho_3$, we have $B_f=1/2$
and $J_{z,f}=0$.

\begin{figure}[t]
\begin{center}
\centerline{
\includegraphics[width=3.0in]{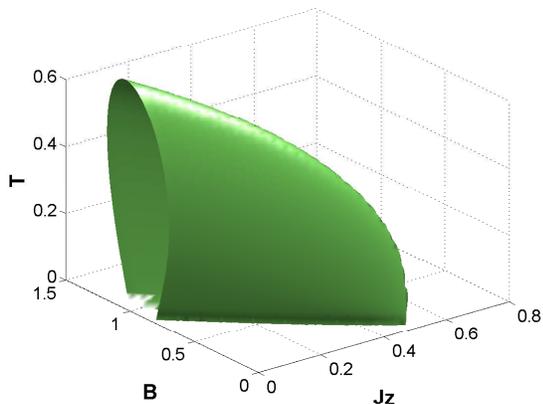} }
\end{center}
\caption{This plots $J_z$ versus $B$ and $T$ when $N=7$, and shows the values
for which we can detect entanglement in $\rho^*$. The state is entangled
between the surface of the plot and the axes.}
\label{fig3}
\end{figure}

We can now detect entanglement in an arbitrary equilibrium state, $\rho^*_3$ using
the entanglement witness. Fig. (\ref{fig2}) shows
the values of the magnetic field, $J_z$ and the temperature for which we
can detect entanglement: we can detect that $\rho^*_3$ is entangled in the
region between the surface and the axes.
Hence, experimentally driving a system from the state $\rho^*_3$ with
values of $B$, $J_z$ and $T$ that are within the surface to the state $\rho$
will allow entanglement to be detected on comparison with the same process
starting at $\sigma_{css,3}$.

Our second example is the $7$ qubit spin chain, with
$B$ time dependent and $J_z=0$. For any $7$ qubit chain, we cannot identify
$\sigma_{css,7}$ with a thermal equilibrium state, and hence we use
$\sigma_7 '=[(7^7-7\times 6^6)|0000000 \rangle \langle 0000000 | + 6^6 \hat{S} (|0000001\rangle)
\hat{S} (\langle 0000001 |)  ]/7^7$ instead. For the initial state to be
$\sigma_7 '$, we require that $B_i= \beta^{-1} \log[70993/46656]/2 + J$
at a low temperature. For concreteness, we again take $\beta^{-1}=0.01$ and
$J=1$, and for the final state to be $\rho_7$, we have $B_f=0.92$.

We can now detect the entanglement of a state $\rho^*_7$ as before. Fig. (\ref{fig3})
shows the values of $B$, $J_z$ and the temperature for which we can detect
entanglement. Again, we can detect that $\rho^*_7$ is entangled in the region
between the surface and the axes.

We note that although $J_z=0$ for both the initial and final state Hamiltonians,
this is not necessarily so for $\rho^*$. Indeed, we can detect when $\rho^*$ is
entangled in many other situations. This is due to the fact that the Hilbert space
of the Hamiltonian is spanned
by the set of $n$ computational eigenvectors, $\{|00 \cdots 0 \rangle ,|00 \cdots 01 \rangle
\cdots |11 \cdots 1 \rangle \}$.
Hence the entanglement witness applies to any state $\rho^*$ that exists within
this Hilbert space.
For example, we could introduce a Dzyaloshinskii-Moriya interaction 
to the Hamiltonian of $\rho^*$ and still use the
witness to detect entanglement in the system.

A possible application of this work is the detection of
entanglement in biological systems. The
photosynthetic bacteria, Prosthecochloris aestuarii, can be
modelled using a seven spin Hamiltonian.
Using experimental values \cite{adolphs,plenio} and
simplifying the model to an isolated system, we can
use this Hamiltonian to construct a specific state $\rho^*$.
The $7$ qubit chain defined above can then be used in the witness. In this
simplified model, we do not detect any entanglement. However,
we expect that the full model, and a more appropriate Hamiltonian
$H_f$ which is closer to $H^*$ will allow entanglement
to be detected.


We have presented a witness which uses
the relative entropy to detect entanglement. When the states are in
equilibrium, we have shown that Jarzynski's equality can be used
to detect entanglement. Hence this witness enables
entanglement to be detected using non-equilibrium processes.
Using this witness, we have considered two examples. In one we can
define an equilibrium closest separable state to $\rho$, and in
the other we instead define an entangled equilibrium state
which has the same directed distance to $\rho$ in terms of the
relative entropy.

\emph{Acknowledgements}: V.V. and J.H. acknowledge the EPSRC for
financial support. V.V. is grateful for funding from the Wolfson
Foundation, the Royal Society and the E.U. His work is
also supported by the National Research Foundation
and the Ministry of Education (Singapore).

\end{document}